\documentstyle[aps,prl,multicol]{revtex}
\def\beginwide{
        \end{multicols} \vspace*{-0.5cm} \noindent
        \rule{3.5in}{.1mm}\rule{.1mm}{5mm} \widetext \medskip }
\def\beginwidetop{
        \end{multicols} \vspace*{-0.5cm} \noindent
        \widetext \medskip }
\def\endwide{
        \hspace*{3.5in}~\rule[-5mm]{.1mm}{5mm}\rule{3.5in}{.1mm}
        \begin{multicols}{2} \vspace*{-1.0cm} \noindent }
\def\endwidebottom{
        \begin{multicols}{2} \vspace*{-1.0cm} \noindent }
\tighten
\draft

\begin{document}
\title{Domain wall dynamics in soft magnetic materials}
\author{Alexei V\'azquez$^1$ and Oscar Sotolongo-Costa$^{1,2}$}
\address{$^1$ Department of Theoretical Physics, Faculty of
Physics, Havana University, Havana 10400, Cuba}
\address{$^2$ LICTDI, Faculty of Sciences, UNED, Madrid 28080,
Spain}
\maketitle

\begin{abstract}
The dynamics of a domain wall in magnetostrictive materials is
investigated.  The domain wall is modeled by a $d$-dimensional
interface moving in a $d+1$-dimensional environment. Long-range
demagnetization effects and quenched disorder are considered,
while the external magnetic field is increased at constant rate.
Exact expressions for the average interface velocity and
susceptibility are obtained, resulting that the system is
critical when the demagnetization constant and the average
interface velocity vanishes. The critical exponents are computed
to ${\cal O}(\varepsilon=4-d)$. Our predictions are compared
with numerical simulations and Barkhausen noise measurements
reported in the literature.
\end{abstract}


\begin{multicols}{2}

Domain walls in ferromagnets move following an irregular motion
in response to changes in an applied external magnetic field,
leading to discrete jumps in the magnetization. This phenomenon,
known as the Barkhausen effect, has received a renewed
theoretical and experimental interest, because of its connection
with non-equilibrium critical phenomena like interface depinning
and self-organized criticality
\cite{urbach,narayan,cizeau,bahiana,durin}.

Urbach {\em et al} \cite{urbach} have analyzed the role of
demagnetization effects in the avalanche statistics. From
numerical simulations and Barkhausen noise measurements, they
conclude that the addition of an "infinite-range"
demagnetization field results in self-organized criticality,
while a local demagnetization field does not. On the other hand,
Narayan \cite{narayan} has shown that long-range dipolar
interactions give rise to self-similar dynamics. More recently,
Cizeau {\em et al} \cite{cizeau} derived an equation of motion
for the dynamics of a ferromagnetic domain wall, including both
long-range demagnetization effects and dipolar interactions.
From numerical simulations, they conclude that long-range
dipolar interactions decreases the upper critical dimension from
$d_{uc}=4$ to $d_{uc}=2$ \cite{note}. Moreover, they have also
demonstrated that the demagnetization factor acts as a control
parameter, criticality is obtained after it is fine tunned to
zero.

In magnetostrictive materials magneto-elastic interactions
enhance the surface tension effect over dipolar interactions
and, therefore, scaling exponents for two-dimensional domain
walls should be lower than the mean-field ones. Experimental
measurements of Barkhausen noise with different magnetostrictive
materials corroborate this conclusion
\cite{urbach,bahiana,durin}.

In a recent work \cite{vazquez} we have analyzed the dynamics of
a $d$-dimensional interface under a driving force increasing
linearly with time, a restored force proportional to the local
interface height $H(\vec{x},t)$, and quenched disorder (This
model is equivalent to the Urbach {\em et al} model with local
demagnetization effects).  Extending previous renormalization
group (RG) calculations developed for the constant force case
\cite{fisher,review} we obtained the dynamic $z$ and roughness
$\zeta$ exponents in first order of $\varepsilon=4-d$. Here we
use a similar approach to analyze the same model but with
long-range demagnetization effects. The starting equation of
motion is a particular case of that obtained by Cizeau {\em et
al} \cite{cizeau}, excluding long-range dipolar interactions.
It is demonstrated that the model with long-range
demagnetization effects belongs to the same universality class
as that of local demagnetization effects \cite{vazquez}. Our
predictions are compared with numerical simulations and
Barkhausen noise measurements, reported in the literature.

In the central part of the hysteresis loop, around the
coercitive field, the magnetization process is mainly due to
domain wall motion, while domain nucleation and rotation are
absent. Moreover in most ferromagnetic materials, due to
magnetocrystalline forces or sample shape, the magnetization has
preferred directions. In the simplest situation, there is a
single easy axis of magnetization and the domains are separated
by surfaces, parallel to the magnetization and spanning the
sample from end to end. In this situation the domain wall
dynamics may be modeled as the motion of a $d$-dimensional
interface in a $d+1$ environment, with height $H(\vec{x},t)$. In
magnetostrictive materials the long range dipolar interactions
can be neglected, resulting the equation of motion \cite{cizeau}
\beginwide
\begin{equation} 
\lambda \frac{\partial}{\partial t} H(\vec{x},t)
=\Gamma\nabla^2H(\vec{x},t)+ht-
\epsilon\int\frac{d^dx^\prime}{L^d}
H(\vec{x}^\prime,t)+\eta[\vec{x},H(\vec{x},t)],
\label{eq:1} 
\end{equation}
\endwide
where $\Gamma$ is the surface tension of the wall, $\lambda$
represents a friction coefficient, $ht$ is the applied field
which is supposed to increase at rate $h$, $L$ is the linear
size of the system, and $\epsilon$ is the demagnetization
factor. Here $\eta(\vec{x},H)$ is a Gaussian uncorrelated noise
due to impurities, with zero mean and
\begin{equation}
\langle\eta(\vec{x},H)\eta(\vec{x}^\prime,H^\prime)\rangle =
\delta^d(\vec{x}-\vec{x}^\prime)\Delta(H-H^{\prime}), 
\label{eq:2} 
\end{equation}
where $\Delta(H)$ is a monotonically decreasing even function,
with a fast decay to zero beyond a distance $a_\bot$. In eq.
(\ref{eq:1}) we have explicitly taken into account that the
demagnetization constant scales as the inverse of the system
size, which for an hyper-cubic lattice scales as $L^d$. Hence,
the demagnetization factor $\epsilon$ does not depend on system
size. Eq. (\ref{eq:1}) is similar to Edwards-Wilkinson equation
with quenched noise, but the external field increases at rate
$h$ and there is an additional term due to demagnetization
effects. This difference carry as a consequence that the
interface is never pinned by impurities, but always moves with a
finite average velocity $v$.

A perturbative solution of eq. (\ref{eq:1}) can thus be found
expanding $H(\vec{x},t)$ around the flat co-moving interface
$vt$. Taking $H(\vec{x},t)=vt+w(\vec{x},t)$, with $\langle
w(\vec{x},t)\rangle=0$, we obtain the following equation for
$w(\vec{x},t)$
\beginwide
\begin{equation} 
\lambda \frac{\partial}{\partial t} w(\vec{x},t)
=\Gamma\nabla^2w(\vec{x},t)+(h-\epsilon v)t-\lambda v
-\epsilon\int\frac{d^dx^\prime}{L^d}
w(\vec{x}^\prime,t)+\eta[\vec{x},vt+w(\vec{x},t)].
\label{eq:4} 
\end{equation}
\endwide
The average velocity is obtained using the constraint $\langle
w(\vec{x},t)\rangle=0$, while fluctuations around the average
will be analyzed using RG transformations. For these purposes is
better to work with the equation for the Fourier transform of
$w(\vec{x},t)$, $\tilde{w}(\vec{k}$,$\omega)$ \cite{review}. The
external field $(h-\epsilon v)t$ gives a singular term of the
order of $\omega^{-2}$. This singular term predominates in the
low frequency limit resulting, after imposing
$\langle\tilde{w}(\vec{k},\omega)\rangle=0$,
\begin{equation}
v=\frac{h}{\epsilon}.
\label{eq:8}
\end{equation}

Another exact result can be obtained if one computes the
low-frequency and long-wavelength susceptibility. Adding a
source term $\hat{\varphi}(\vec{k},\omega)$ to the equation for
$\tilde{w}(\vec{k}$,$\omega)$ one obtains the generalized
response function \cite{review}
\begin{equation}
\tilde{G}(\vec{k},\omega)=\frac{\tilde{w}(\vec{k},\omega)}
{\tilde{\varphi}(\vec{k},\omega)}\bigg|_{\tilde{\varphi}=0}=
\frac{1}{[\tilde{G}_0(\vec{k},\omega)]^{-1}-
\tilde{\Sigma}(\vec{k},\omega)},
\label{eq:9}
\end{equation}
where
\begin{equation}
[\tilde{G}_0(\vec{k},\omega)]^{-1}=\Gamma\vec{k}^2+
i\lambda\omega+\epsilon\hat{\delta}(\vec{k},L).
\label{eq:6}
\end{equation}
is the bare correlator and $\tilde{\Sigma}(k,\omega)$ is the
"self-energy". $\hat{\delta}(\vec{k},L)$ is the Fourier
transform of $L^{-d}$. In the thermodynamic limit
($L\rightarrow\infty$) $\hat{\delta}(\vec{k},L)\approx1$ for
$\vec{k}\rightarrow0$ and zero otherwise. Since
$\tilde{\Sigma}(0,0)=0$ and $\tilde{G}_0(0,0)=\epsilon^{-1}$ it
results that the low-frequency and long-wavelength
susceptibility (or simply the susceptibility) is given by
\begin{equation}
\chi=\tilde{G}(0,0)=\frac{1}{\epsilon}.
\label{eq:10}
\end{equation}
This result is also exact.

In the limit $\epsilon\rightarrow0$ the susceptibility diverges
as $\chi\sim\epsilon^{-\gamma}$ with
\begin{equation}
\gamma=1.
\label{eq:11}
\end{equation}
The system is thus critical for $\epsilon\rightarrow0$, i.e.
when demagnetization effects becomes negligible. However,
different behaviors can be observed depending on the value of
$h$. If $h$ is finite and we take $\epsilon\rightarrow0$ then,
according to eq. (\ref{eq:8}), the domain wall velocity will
become infinitely large and, therefore, the noise term in eq.
(\ref{eq:4}) will be reduced to an annealed noise. Thus, for
($h>0$, $\epsilon\rightarrow0$) the interface dynamics is
described by the EW equation with annealed noise, which has an
upper critical dimension 2. Hence, for $d\geq2$ the interface
will moves with a velocity given by eq. (\ref{eq:8}) and
fluctuations around the average can be neglected. This picture
corresponds with a suppercritical regime, where the external
magnetic field predominates over disorder. On the contrary, for
($h\rightarrow0$, $\epsilon>0$) the system will be in a
subcritical regime. The domain wall will moves with an
infinitely small velocity and, therefore, the quenched noise
will lead to fluctuations around the flat co-moving interface,
but these fluctuations will be not critical due to
demagnetization effects. This fact becomes evident with the
existence of a finite susceptibility. Finally, the critical
state corresponds with the double limit $\epsilon\rightarrow0$
and $v=h/\epsilon\rightarrow0$.

To go further we perform a RG analysis of the problem. We
integrate out the degrees of freedom in a momentum shell near
the cutoff $\Lambda$ and rescale $k\rightarrow b^{-1}k$,
$\omega\rightarrow b^{-z}\omega$, and $\tilde{w}\rightarrow
b^{\zeta+d+z}\tilde{w}$, where $b=\text{e}^l$ with
$l\rightarrow0$. The flow equations for the parameters $\Gamma$,
$\lambda$, $\epsilon$ and $v$ are obtained through a direct
application of the RG transformations to eq. (\ref{eq:4}) in the
Fourier space resulting, to one-loop order,
\begin{equation}
\frac{d\Gamma}{dl}=0,\ \ \ \
\frac{d\lambda}{dl}=\lambda(2-z+cQ_2),
\label{eq:13}
\end{equation}
\begin{equation}
\frac{d\epsilon}{dl}=2\epsilon,\ \ \ \ 
\frac{dv}{dl}=(z-\zeta)v.
\label{eq:15}
\end{equation}
where $c=\frac{S_d}{(2\pi)^d\Gamma^2}\Lambda^{-\varepsilon}$,
$S_d$ is the complete solid angle in $d$ dimensions, and
$Q_2=\int_q\tilde{\Delta}(q)q^2$ is the second moment of the
noise correlator. Here we have implicitly assumed that
$\Gamma\Lambda^2\gg\epsilon$ and $\Gamma\Lambda^2\gg\lambda
v/a_\bot$, as it is expected near the critical point where both
$\epsilon$ and $v$ should vanishes. The long-range nature of
demagnetization effects is, within our RG approach, irrelevant.
The same result will be obtained if one considers a local
demagnetization term $-\epsilon H$. Hence, all the analysis
developed below will also hold for local demagnetization
effects.

The renormalization of the moments of the noise correlator
($Q_{n}=\int_q\tilde{\Delta}(q)q^n$) is obtained considering a
vertex function \cite{review,fisher}. The renormalization
equations for the moments can be put all together in a
renormalization equation for the noise correlator $\Delta(H)$,
obtaining
\beginwide
\begin{equation}
\frac{\partial\Delta(H)}{\partial l}=(\varepsilon-2\zeta)\Delta(H)+
\zeta H\Delta^\prime(H)-c\frac{d^2}{dH^2}\left[
\frac{1}{2}\Delta^2(H)-\Delta(H)\Delta(0)\right].
\label{eq:18}
\end{equation}
\endwide
We look for a fixed point solution $\Delta^*(H)$ to this
equation, obtained setting $\partial\Delta/\partial l$ to zero
and adjusting $\zeta$. Integrating from $-\infty$ to $\infty$
one obtains \cite{fisher,review}
\begin{equation}
\zeta=\frac{\varepsilon}{3},
\label{eq:19}
\end{equation}
provided $\int_{-\infty}^\infty dH\Delta^*(H)\neq0$. On the
other hand, for $H\rightarrow0$ we obtain $\Delta^*(H)\approx
Q_0+Q_1|H|-\frac{1}{2}Q_2H^2$ with
\begin{equation}
cQ_2=-\frac{\varepsilon-\zeta}{3}.
\label{eq:20}
\end{equation}
Then, imposing scale invariance in eq. (\ref{eq:15}) and using
eqs. (\ref{eq:19}) and (\ref{eq:20}) it results that
\begin{equation}
z=2-\frac{2}{9}\varepsilon.
\label{eq:21}
\end{equation}
The exponents $\zeta$ and $z$ thus result identical to those
obtained for the constant force case \cite{fisher,review}.

On the other hand, eq. (\ref{eq:15}) implies that there is a
characteristic lenght $\xi\sim\epsilon^{-\nu}$ beyond which
scale invariance is lost due to demagnatization effects, and a
characteristic velocity $v_c\sim\epsilon^\theta$ (or field rate
$h_c=\epsilon v_c$) beyond which the system is supercritical,
where
\begin{equation}
\nu=\frac{1}{2},\ \ \ \ \theta=\nu(z-\zeta).
\label{eq:22}
\end{equation}
The line $v_c\sim\epsilon^\theta$ devides the phase diagram
$(v,\epsilon)$ into two regions. For $v>v_c$ the noise is
annealed and, therefore, the model is supercritical as discussed
above. On the contrary, for $v<v_c$ the model is subcritical for
all $\epsilon>0$. The critical state is obtained in the double
limit $\epsilon\rightarrow0$ and $v=h/\epsilon\rightarrow0$. The
scaling exponents $\gamma=1$ and $\nu=1/2$ are exact while
$\zeta$ and $z$ are given by eqs. (\ref{eq:19}) and
(\ref{eq:21}), respectively, up to the first order in
$\varepsilon$.

In the subcritical regime the dynamics takes place in the form
of avalanches, characterized by the avalanche size
$P(s)=s^{-\tau}f(s/s_c)$ and duration $P(T)=T^{-\alpha}g(T/T_c)$
distributions, where $s_c$ and $T_c$ are the avalanche size and
duration cutoffs. In the subcritical state
$s_c\sim\epsilon^{-1/\sigma}$. At criticality $s_c\sim L^D$ and
$\xi\sim L$, where $D$ is the avalanche dimension, leading to
the scaling relation $\sigma=1/D\nu$.  Other scaling relations
are obtained taking into account that $\chi=\langle s\rangle$
and $\int dsP(s)=\int dTP(T)$, which lead to
$\gamma=(2-\tau)/\sigma$ and $(\tau-1)D=(\alpha-1)z$,
respectively. Taking $\gamma$, $\nu$ and $D$ as independent
exponents we obtain the scaling relations for the avalanche
exponents
\begin{equation}
\tau=2-\frac{\gamma}{D\nu},\ \ \ \ 
\alpha=1+\frac{D-\gamma\nu^{-1}}{z}.
\label{eq:23}
\end{equation}
On the other hand, for $d<d_c$, the avalanche dimension and the
roughness exponent are related via $D=d+\zeta$. Hence, we can
compute $\tau$ and $\alpha$ using the RG estimates of $\zeta$
and $z$ and the exact values of $\gamma $ and $\nu$. Above the
upper critical dimension $d_{uc}=4$, one obtain the mean-field
exponents $D=d_{uc}=4$, $\tau=3/2$ and $\alpha=2$.

Before compare our exponent predictions with experiments and
simulations, let us make some remarks about experimental
measurements. The Barkhausen signal $V(t)$ is the voltage
produced from a pickup coil around a ferromagnet subjected to a
slowly varying applied field. In the low-frequency limit the
time scale for domain wall motion is much smaller than the time
between jumps and, therefore, one may guarantee that each
induced voltage jump corresponds with a single avalanche in the
domain wall motion. A resolution voltage level $V_R$ is defined,
such that one can not resolve details below $V_R$. An elementary
Barkhausen jump can thus be defined as the portion of the $V(t)$
signal delimited by two subsequent intersections of the signal
with the $V_R$ line. With this definition, the duration $T$ is
simply the time interval between these two subsequent
intersections and the size $s$ is the area delimited by $V(t)$
and $V_R$ between the same points.

Evaluating the scaling exponents for $d=2$, using the scaling
relations in eqs. (\ref{eq:23}), the RG estimates of $\zeta$ and
$z$ and the exact values of $\gamma$ and $\nu$,  we obtain
$\tau=5/4=1.25$ and $\alpha=10/7\approx1.43$.  These values can
be compared with Barkhausen measurements in magnetostrictive
materials \cite{urbach,bahiana,durin}. Earlier measurements by
Urbach {\em et al} \cite{urbach} gives $\tau=1.33\pm0.10$. More
recently, Durin and Zapperi (DZ) \cite{durin} reported the more
accurate exponents $\tau=1.28\pm0.02$ and $\alpha=1.5\pm0.1$,
which are close to our estimates. In the last case the
experiments were performed under an applied stress. Positive
magnetostrictive materials show an increase in internal domain
wall lengths under applied stress, which make possible the
analysis of finite size effects in experimental results.
According to DZ measurements, the signal amplitude cutoff
increases with the stress while the avalanche duration cutoff
decreases with stress, in such a way that the avalanche size
cutoff $s_c$ is nearly independent on stress. The stress
dependence of the signal amplitude cutoff was also observed in
\cite{bahiana}. We still do not  understand the scaling of the
distribution cutoffs with applied stress.  However, the absence
of stress dependence in $s_c$ suggests that the system is in a
subcritical state, where the correlation length is smaller than
system size. Unfortunately, the demagnetization factor is kept
constant in these experiments and, therefore, one can not
determine if it is actually a control parameter.

On the other hand, numerical simulations of the cellular
automaton version of eq. (\ref{eq:1}) have also been performed
\cite{bahiana,durin}. The more accurate numerical estimates,
reported by Durin and Zapperi \cite{durin}, are
$\tau=1.26\pm0.04$ and $\alpha=1.40\pm0.05$, which are in very
good agreement with our estimates. However, as in experiments,
the demagnetization factor is kept constant. Numerical
simulations of the same model, but including long-range dipolar
interactions, clearly shows that the demagnetization factor is a
control parameter \cite{cizeau}. We expect the same result when
long-range dipolar interactions are absent.

Hence, we strongly suggest to change the demagnetization factor
in both, experiments and simulations. In numerical simulations
it is reduced to change a model parameter, while it can be done
in experiments changing, for instance, sample geometry. Changing
the demagnetization factor will be a very good test to our model
predictions. After that, one may analyze if the avalanche
distribution cutoffs are a consequence of finite size effects or
just an evidence that the system is in a subcritical state.
Based on numerical simulations, Bahiana {\em et al}
\cite{bahiana} rule out the second possibility. However, they
may have taken the demagnetization factor so small that the
correlation length is smaller than system size. Decreasing the
demagnetization factor the system will, according to our
predictions, evolve to the subcritical state.

We conclude that the long-range demagnetization field does not
change the universality class from that of local demagnetization
field. This result contradicts previous conclusion by Urbach
{\em et al} \cite{urbach}, based on numerical simulations in one
and two dimensions and Barkhausen noise measurements. According
to them, the addition of a long-range demagnetization field
results in self-organized criticality, while a local
demagnetization field does not. However, in our opinion, their
analysis only reveals that long-range demagnetization effects
are more appropriate to describe the domain wall motion in
magnetic materials. As in other numerical simulations and
experiments, the demagnetization factor is kept constant.

This work has been partly supported by the {\em Alma Mater }
prize , from the Havana University, and Ministerio de
Educaci\'on y Cultura, Spain.

\end{multicols}


\begin{thebibliography}{50}
\bibitem{note} Here $d$ denotes the domain wall dimension, which
moves in a $d+1$-dimensional enviroment.
\end{thebibliography}
\end{document}